\documentstyle[11pt]{article}

\def\pr#1#2#3{{\elevenit   Phys. Rev.} ${\bf{#1}}$ (#2) #3 }
\def\prl#1#2#3{{\elevenit  Phys. Rev. Lett.} ${\bf{#1}}$ (#2) #3 }
\def\pl#1#2#3{{\elevenit   Phys. Lett.} ${\bf{#1}}$ (#2) #3 }

\def\np#1#2#3{{\elevenit   Nucl. Phys.} ${\bf{#1}}$ (#2) #3 }
\def\zp#1#2#3{{\elevenit   Z. Phys.} ${\bf{#1}}$ (#2) #3 }
\def\ijmp#1#2#3 {{\em Int. J. Mod. Phys.} {\bf#1} (#2) #3}

\def\cost{\cos\theta}
\def\sint{\sin\theta}
\def\cop{\cos\varphi}
\def\sip{\sin\varphi}
\def\su{\sin\theta}

\def\mw{m_W^2}
\def\mz{m_Z^2}
\def\mv{m_V^2}

\def\p{\vert\vec p \, \vert}
\def\pp{\vert\vec p\, ' \vert}
\def\rs{\sqrt{s}}

\def \b {\hskip -5pt/}

\def\l{{\cal L}}
\def\v{{\cal V}_\mu}
\def\ep{\epsilon}
\def\epp{\epsilon^\prime}
\def\ctw{\cot\theta_W}
\def\tw{\theta_W}
\def\g{\gamma}
\def\epst{\epsilon_{\tau\nu\rho\sigma}}
\def\eps{\epsilon_{\mu\nu\rho\sigma}}
\def\eph{\epsilon^{\mu\nu\rho\sigma}}
\def\epj{\epsilon^{\sigma\rho\alpha\beta}}
\def\vb{\bar{v}}

\def\ds{{d\sigma\over d\cos\theta}}

\def\n{{\cal N}}

\def\f{{\cal F}}
\def\G{{\cal G}}

\def\pa{\partial}

\def\o{{\cal O}}

\def\ee{e^+e^-\to W^+W^-}

\def\bq{\begin{equation}}
\def\eq{\end{equation}}
\def\bqa{\begin{eqnarray}}
\def\eqa{\end{eqnarray}}
\def\bqb{\begin{eqnarray*}}
\def\eqb{\end{eqnarray*}}

\def\ww{\bf q^{(')}_i \bar{q}^{(')}_i  \rightarrow W^+W^-}
\def\wz{\bf q_i       \bar{q}'_j       \rightarrow W^\pm Z}
\def\wg{\bf q_i       \bar{q}'_j       \rightarrow W^\pm \gamma}
\def\zz{\bf q^{(')}_i \bar{q}^{(')}_i  \rightarrow ZZ}
\def\zg{\bf q^{(')}_i \bar{q}^{(')}_i  \rightarrow Z\gamma}
\def\gg{\bf q^{(')}_i \bar{q}^{(')}_i  \rightarrow \gamma\gamma}

\def\bs{\beta^2}
\def\chiw{\chi_{\raise -1.5mm\hbox{\footnotesize\it W}}}
\def\chiz{\chi_{\raise -1.5mm\hbox{\footnotesize\it Z}}}

\def\avi{a_{Vi}}
\def\avj{a_{Vj}}
\def\bvi{b_{Vi}}
\def\bvj{b_{Vj}}
\def\dps{\displaystyle}
\def\ni{\noindent}
\def\dag{\dagger}

\font\elevenbf=cmbx10 scaled\magstep 1
\font\elevenrm=cmr10  scaled\magstep 1
\font\elevenit=cmti10 scaled\magstep 1
\renewenvironment{thebibliography}[1]
 { \elevenrm
   \begin{list}{\arabic{enumi}.}
    {\usecounter{enumi} \setlength{\parsep}{0pt}
     \setlength{\itemsep}{3pt} \settowidth{\labelwidth}{#1.}
     \sloppy
    }}{\end{list}}

\begin{document}
\hoffset=-0.5truecm
\voffset=+0.8truecm
\textheight 22cm
\pagenumbering{arabic}
\thispagestyle{empty}
\hspace {-0.8cm} PM/96-30\\
\hspace {-0.8cm} October 1996\\
\vspace {0.8cm}\\
\vspace{1cm}
\begin{center}
{\Large\bf  Di-Boson production at Hadron Colliders with }\\
\vspace {0.1cm}
{\Large\bf  General 3 Gauge Boson Couplings.}\\ 
\vspace {0.1cm}
{\Large\bf  Analytic Expressions of Helicity Amplitudes and 
            Cross-Section. }\\

\vspace{1.8cm}
{\large  E. NUSS}
\vspace {1cm}  \\
Physique Math\'{e}matique et Th\'{e}orique,
CNRS-URA 768,\\
Universit\'{e} de Montpellier II,
 F-34095 Montpellier Cedex 5.\\
\vspace{1.5cm}

{\bf Abstract}
\end{center}
\noindent
We discuss the tests of general Three Gauge Boson Vertices (TGV) through
 bosonic pair production at present and future hadron colliders.
All bosonic final states are reviewed via the tree level quark-antiquark 
annihilation sub-process.
The full analytic expressions of the helicity amplitudes and 
cross-sections are given.
These expressions should be useful in any attempt to disentangle the 
effects of the most general non standard $WWV (V=\gamma,Z)$ vertices 
including 14 free parameters.
We investigate the sensitivity of the invariant mass and transverse 
momentum distributions to the full set of anomalous couplings including 
final state polarization structures.
We particularly consider these features at the projected CERN Large 
Hadron Collider (LHC) energy scale.

\newpage
\vglue 0.6cm
{\elevenbf\noindent 1. Introduction.}
\vglue 0.2cm
\baselineskip=14pt
\elevenrm

It is well known that in spite of its elegant theoretical construction 
and a lot of experimental confirmations, the Standard Model (SM) cannot 
be considered as the final theory of matter and its interactions.
As it can be done for the Fermi theory in regard of the SM, it seems 
reasonable to consider the electroweak theory as the low energy residual 
part of a more fundamental theory supposed to be found at higher energy 
scale.
It is clear that the crucial point of this hypothesis is to find a direct
 or indirect signal of this "new" physics (NP).
Numerous theoretical studies strongly suggest that NP is to be expected 
in the TeV range.

\ni
The high accuracy of experiments on electroweak measurements at LEPI and 
SLC probe the predictions of the SM at few $\times 10^{-3}$ 
level\cite{smtest} and do not lead to any undisputable deviation albeit 
possible deviations in $A_b$ \cite{Ab}.
With the same proviso on the measurement of heavy quarks observables, 
no departure from the SM is measured up to FERMILAB energy scale 
\cite{smtestHADCOL}.
The task of future experiments will be to probe the physics beyond SM. 
If NP scale is $\geq 1$ TeV, it is not certain that future experiments 
will find it first directly e.g. through the discovery of new interacting 
particles and we have to look for indirect experimental evidences.
As it is briefly reviewed below, the bosonic sector remains poorly tested
 and, if no Higgs particles is discovered up to the $1$ TeV scale, would 
become one of the most privileged sector for exploring NP like composite 
$W$'s, strongly interacting longitudinal $W$'s or virtual effects of new 
heavy particles.

\ni
At present colliders the process of gauge boson pair production is just 
below the kinematical limits and the low rate of events leads to very 
large errors.
Given the actual measurements, the bosonic self interactions remain 
allowed to deviate considerably from their standard expectations but 
the construction in a near future of a new generation of accelerators 
will allow us for the first time to make some direct and precise 
measurements of all three and four boson couplings predicted by the 
standard electroweak theory.
In the SM, these W,Z, and photon self-couplings are strongly related to 
the scalar sector through the longitudinal components of the vector 
bosons \cite{wl}.
Thus, they are an open window on the electroweak gauge symmetry breaking 
mechanism.
They are also a direct manifestation of the non-abelian underlying 
$SU(2)_L \otimes U(1)_Y$ gauge structure \cite{sm} on which the theory 
is based. 
In the case of NP related to the mass generation sector, to the gauge 
symmetry or either to a compositeness of weak-bosons, it is natural to 
expect that various kinds of anomalous trilinear or quartic couplings 
will be generated \cite{NPandAC}. 
In the next section we will recall how they can be modelized by some 
effective anomalous couplings. 
A direct experimental search for evidence of such anomalous couplings 
appears clearly as a fundamental test to reach this kind of NP.

\ni
As we know, the unitarity of the SM depends directly on its gauge 
structure. 
Important cancellations between the amplitudes participating to a process
 ensure the decrease with energy of the integrated cross-section.
Any departure from this vector bosonic gauge structure would break the
 unitarity at low energy which should leads to an indirect signal of NP.
As energy scale is increased, the unitarity should be restored by the
 contribution of new degrees of freedom.

\ni
The main problem is firstly, to test NP existence through the measurement
 of anomalous couplings.
Secondly, we have to characterize this possible NP by an identification
 of the effective couplings generating those deviations.

\vspace{0.5cm}
Because they are experimentally more accessible and easily treated by 
effective theories we restrict ourself to the Three Gauge Boson 
Vertices (TGV)
\footnote{For tests of quartic couplings in colliders experiments, see 
Ref.\cite{quartic} }.
Four Gauge Boson Vertices could be related to the previous one in a 
scenario where the anomalous couplings are generated by the symmetry 
breakdown of gauge invariant operators \cite{hagiOpe}.

\vspace{0.5cm}
This paper is organized as follows. 
In Sec.2 we recall some definitions and settle our notations for the 
general $WWV$ ($V=\gamma,Z$) effective Lagrangian. 
We briefly link this effective approach to the description of anomalous 
couplings in term of gauge invariant operators. 
In Sec.3, we give the explicit expression of the invariant mass and 
transverse momentum distributions in terms of the usual partonic 
description used in our numerical computations. 
In Sec.4, we describe the procedure we apply to obtain the helicity 
Tables for di-boson production sub-processes which are computed 
analytically in Sec.5 and 6 for 
$W^+ W^- , W^\pm Z$ and $W^\pm \gamma$ productions. 
For completeness, we deduce the helicity Tables for the standard $ZZ$,
 $Z\gamma$ and $\gamma\gamma$ production from the previous one.
We give also the resulting analytic expressions of sub-process 
cross-section in function of the complete set of anomalous free 
parameters. 
In Sec.7, we discuss these results and the implication of the various
 anomalous couplings to the hadronic observables.
We compare standard and non standard shapes of invariant mass and 
transverse momentum distribution for LHC energy scale. 
We also give the shapes of the anomalous coupling contributions to the 
final bosonic polarization states allowed for each productions. 
Using the fact that the various anomalous couplings contribute to 
helicity amplitudes in different ways, we suggest a possible procedure
 to be applied in order to disentangle these couplings.
\vspace{0.5cm}

\vglue 0.6cm
{\elevenbf\noindent 2. Theoretical Background}
\vglue 0.2cm
\baselineskip=14pt
\elevenrm

In the absence of specific models, the effective Lagrangian approach 
\cite{lageff} is extremely useful to parameterize in a model independent
 way the low-energy effects of NP.
We recall that in the approximation of bosons coupled to massless 
fermions and for on-shell final vector bosons, the most general Lorentz
 invariant structure for $W^+W^-V$ trilinear vertex
\footnote{ Analogously to the general WWV vertex it is possible to 
parameterize anomalous $Z\gamma V,V=Z,\gamma$ couplings \cite{ZGVvertex}
 but these will not be considered here.}
, where $V=Z$ or $\gamma$, involves only seven independent terms. 
This was firstly written in Ref.\cite{lag} and reads
\bqa
\l_{WWV}&=&ie(g^{SM}_{WWV}+\delta_V) 
\Big[ V_\mu(W^{\dag\mu\nu}W_\nu-W^{\mu\nu}W^
     \dag_\nu)+V_{\mu\nu}    W^{\mu}W^{\dag\nu} \Big] \nonumber\\
& &\mbox{}+iex_V V^{\mu\nu}W_\mu
W^\dag_\nu+{iey_V\over\mw}V^{\nu\lambda}W_{\lambda\mu}
W^{\dag\mu}_{\mbox{  }\ \ \nu} \nonumber\\
& &\mbox{}+{ez_V\over\mw}(\pa^\alpha\tilde V^{\rho\sigma})
\Big\{ (\pa_\rho W_\sigma)W^\dag_\alpha-(\pa_\rho W_\alpha)W^\dag_\sigma+
(\pa_\rho W^\dag_\sigma)W_\alpha-(\pa_\rho W^\dag_\alpha)W_\sigma \Big\} 
\nonumber\\
& &\mbox{}+ez'_{1V}\pa^\rho V^\sigma(W^\dag_\rho W_\sigma+W^\dag_\sigma
W_\rho) +iez'_{2V}\eph\pa_\mu V_\nu W^\dag_\rho W_\sigma \nonumber\\
& & \mbox{}+2ie{z'_{3V}\over\mw}\epj\pa_\sigma V_\mu(\pa^\mu
W_\beta\pa_\rho W^\dag_\alpha+\pa^\mu W^\dag_\alpha \pa_\rho W_\beta)
\label{lag}
\eqa
in the notation of Ref.\cite{renee}.
In Eq.(\ref{lag}), $\ep^{0123}=1$, 
$V^{\mu\nu}=\pa^\mu V^\nu-\pa^\nu V^\mu$ and
$ \widetilde V_{\rho\sigma}=
{1\over2}\ep_{\rho\sigma\alpha\beta} V^{\alpha\beta}\ $
denotes the dual field tensor. 
Higher derivative terms come as form factors of these couplings.\\
The effective Lagrangian (\ref{lag}) is parameterized in terms of 14 free
 parameters.
For each $V=\gamma$ or $Z$ bosons; $\delta_V, x_V, y_V, z_V, z'_{1V},
 z'_{2V}$ and $z'_{3V}$ represent the deviations of the various couplings
 from their $SU(2)_L \otimes U(1)_Y$ standard values.
Electromagnetic gauge invariance fixes to zero $\delta_\gamma$ and 
$z_\gamma$ in the limit of on-shell photon ($q^2=0$). 
The tree level standard case corresponds to $g^{SM}_{WW\g}= 1$ and 
$g^{SM}_{WWZ} =\ctw$ with all anomalous couplings vanishing.
The full description of the different couplings and their relations with
 the equivalent parameterization of Hagiwara et al. \cite{lag} can be 
found in Ref.\cite{renee}.\\
In the SM case, all terms of Eq.(\ref{lag}) are generated by perturbative
 calculations \cite{SMperturb}.
The MSSM contribution to the $x_V$ and $y_V$ terms, related to the usual
 magnetic and quadrupole moments $\kappa_V$ and $\lambda_V$ through 
relations $x_V=(\kappa_V-1)g_V$ and $y_V=\lambda_Vg^{SM}_{WWV}$ where 
$g_V=g^{SM}_{WWV}+\delta_V$, are computed in Ref.\cite{kaplamMSSM} and 
references therein.
The minimal effects in the unconstrained MSSM have been found to be of 
the order of $\Delta\kappa_\gamma=1-\kappa_\gamma \simeq 1.75\ 10^{-2}$
 and $\Delta\kappa_Z=1-\kappa_Z \simeq 0.84\ 10^{-2}$ at LEPII 
($\sqrt{s}=190$ GeV) as compared to a sensitivity of about $5.4\ 10^{-2}$. 
The $\lambda_V$ contributions are about a factor $2-3$ smaller and the 
SM virtual contributions are of the order of 
$4.131\ 10^{-3}-5.505\ 10^{-3}$ for $\Delta\kappa_\gamma$ and 
$3.323\ 10^{-3}-3.148\ 10^{-3}$ for $\Delta\kappa_Z$. 
At NLC ($\sqrt{s}=500$ GeV), the MSSM effects can reach 
$\Delta\kappa_V\simeq 5.4\ 10^{-3}$ with a precision of $0.81\ 10^{-3}$.
In the general case of non-standard physics, several of these couplings
 can simultaneously appear, depending on the underlying dynamics 
\cite{rendyn} and there is no strong theoretical assumptions to privilege
 any one of them. 
In this case, the question of finding a strategy to identify and 
disentangle the effects of the full set of the 14 parameters is raised 
and will be discussed in Sect.7.\\
Let us discuss, how the non $SU(2)_L \otimes U(1)_Y$ gauge invariant 
effective Lagrangian (\ref{lag}) can be induced by gauge invariant 
operators \cite{OpeCA,OpeCA2}.
In this kind of approach, we assume that the NP is invariant under the 
usual local $SU(2)_L \otimes U(1)_Y$ symmetry and that this symmetry is 
spontaneously broken by the vacuum expectation value of the Higgs doublet
 field $\Phi$.    
Integrating out the heavy degrees of freedom, the interactions between 
the gauge bosons and the Higgs doublet lead to an effective Lagrangian 
which describes the residual effects of the full theory at low energy. 
The full Lagrangian may be written \cite{buchmuller}

\bq
\l=\l_{SM}+\sum_{i,d} {f_i \over \Lambda^{d-4}} O^{(d)}_i (x)
\label{efflag}
\eq

\ni
where $O^{(d)}_i$ are local operators with mass dimension $d$ greater 
than four, $SU(2)_L \otimes U(1)_Y$ gauge invariant and involving 
$W$,$Z$,$\gamma$ and Higgs fields only.
The index $i$ runs over all possible operators of a given mass dimension
 $d$.
$\Lambda$ denotes the scale at which NP gives a strong contribution in
 the weak-boson sector. 
According to Eq.(\ref{efflag}) we see that operators of high dimension
 are suppressed by negative powers of $\Lambda$.
An exhaustive list of mass-dimension six operators has been compiled in
 Ref.\cite{hagiOpe} and the explicit form of operators generating the
 full set of TGV's anomalous couplings can be found in Ref.\cite{OpeCA}.

\ni
Several works on TGV study at LEPII through the $\ee$ channel 
\cite{renee,lep} show that despite a relatively clean environment, 
the weak number of events and the 14 parameters contribution to the same
 process will not allow to disentangle the different contributions of 
Eq.(\ref{lag}). 
However, such an identification is crucial in selecting the operators 
generating the anomalous couplings. 
This classification gives fundamental informations on the NP 
\cite{rendyn} as e.g. an idea of its underlying dynamics, energy scale 
and a mass bound for its lowest degrees of freedom.

\ni
As the unitarity is broken, any deviation from SM couplings leads to 
amplitudes growing with energy \cite{violunit,factform}, therefore they 
are more apparent at higher invariant masses of the gauge boson pair but
 the $200$ GeV of LEPII will only allow us to reach the NP scale 
$\Lambda$ of Eq.(\ref{efflag}) below the $\simeq 2$ TeV range.
The future CERN Large Hadron Collider \cite{LHCmachine} [(LHC), pp 
collisions at $\sqrt{S}=14$ TeV] will allow us to explore higher NP 
scales. 
Assuming an LHC year of $\int {\cal L} dt=10^5 \ pb^{-1}$ leading to 
a high rate of production, a detailed study of $VV'$ pair production 
where $V,V'=W^\pm,Z$ or $\gamma$, is particularly interesting to perform
 TGV measurement and separate their effects.
\vspace{0.5cm}

Hadrons colliders are able to produce gauge boson pairs in both charged 
and neutral final states but only the $W^+W^-$, $W^\pm\gamma$ and 
$W^\pm Z$ channels give a TGV contribution.
The first channel suffers from a large QCD background and, as in $\ee$ 
process, is sensitive to both the $WW\gamma$ and $WWZ$ couplings via the
 S-channel. 
In contrast, the $W^\pm\gamma$ and $W^\pm Z$ channels are particularly 
interesting since they are relatively background free and easy to isolate
 compared to $W^+W^-$ pair production \cite{baur,lhc,ampzero}. 
These channels are also particularly interesting since they allow 
independent tests of $WW\gamma$ and $WWZ$ vertices.

\ni
The different contributions of the various terms in Eq.(\ref{lag}) to 
the helicity amplitudes will allow us to disentangle their effects.
The best way to distinguish the anomalous couplings is to identify the
 polarization states of the final bosons for each process which would 
imply the measurement of angular variables in final leptonic decay.\\ 
The $ZZ,Z\gamma$ and $\gamma\gamma$ reactions are not sensitive to the 
previous TGV but could test the structure functions or a possible 
non-standard interaction for example related to compositeness of the 
$Z$ gauge boson (see previous footnote).
\vspace{0.5cm}

\ni
While the contribution of any fixed anomalous coupling $\Delta^0_{g_{V}}$
 of Eq.(\ref{lag}) rises without limit as the sub-process energy 
$\sqrt{\hat s}$ increases, eventually violating partial wave 
unitarity\cite{unitviol2}, we can choose to parameterize their energetic 
behavior like a nucleon form factor \cite{factform}:
$$\Delta_{g_V}(\hat s)=
{\Delta^0_{g_{V}} \over {\dps \Big( 1+{\hat s\over 
\Lambda_{FF}^2} \Big)^n}}$$
where $\Lambda_{FF}$ is a form factor scale depending on the NP scale 
$\Lambda$ and is chosen with $n$ as the minimal value compatible with 
unitarity. 

\ni 
The energy-dependent form factor behavior of anomalous couplings extends 
the use of effective Lagrangians to the entire energy range which is 
accessible at hadron colliders but are based on ad-hoc assumptions by the
 choice of the values of $n$ and $\Lambda_{FF}$.
In the case of high production rate, one can suppress these hypotheses on
 the underlying NP by fitting the couplings assuming they are $\hat s$ 
independent within small energy bins.
\vspace{0.5cm}

\vglue 0.6cm
{\elevenbf\noindent 3. Partonic Description  }
\vglue 0.2cm
\baselineskip=14pt
\elevenrm

We will consider here mainly proton-proton collisions at the LHC energy
 scale ($\sqrt{S}=14$ TeV). 
The results can be easily extended to the $p\bar p$ case with appropriate
 modifications of the structure functions.

\ni
The leading lowest-order processes for di-boson production in $pp$ 
collisions are the boson-boson fusion and the quark-antiquark annihilation
 illustrated in Fig.1.a and Fig.1.b.
In this paper, we will only study the complete case of the Drell-Yann 
mechanisms of quark annihilation and leave the treatment of boson fusion
 for a separate analysis \cite{bosfus}.

\ni
At high energy, these processes are fully described in the partonic 
approximation \cite{partappro}. 
The cross-section $\sigma(pp \rightarrow VV')$ with $V,V'=W^\pm,Z$  or
 $\gamma$ is schematically given by \cite{partons}
\bqa
\dps{ d\sigma= \sum_{ij}
\int \int dx_a \ dx_b \left\{ f_i^{(a)}(x_a)f_j^{(b)}(x_b)+f_j^{(a)} 
(x_a)f_i^{(b)}(x_b) \right\} \ d\hat\sigma_{ij} }
\label{partons02}
\eqa
where $f_i^{(a)}$ are the structure functions and contain informations 
about i-quark luminosity in hadron a. 
$\hat \sigma(q_i\bar q_j'\rightarrow VV')$ is the cross-section for the 
sub-process leading to the desired $VV'$ final state. 
The $x_{i}$ and $x_{j}$ are the momenta fractions of the $(i,j)$-partons
 in the nucleons.

\ni
The ${i,j}$-summation runs over all contributing sea and valence quark
 configurations and depends on the $VV'$ final state. 
Neglecting the top quark contribution we write

\ni
$(ij)=(u \bar u),(d\bar d),(s\bar s),(c\bar c),(b\bar b)$ 
for $W^+W^-,Z\gamma,\gamma\gamma$ and $ZZ$ productions;

\ni
$(ij)=(u\bar d),(c\bar s)$   for $W^+ Z$ or  $W^+\gamma$ productions 
and 

\ni
$(ij)=(\bar u d),(\bar c s)$ for $W^- Z$ or  $W^-\gamma$ final states.

\ni
For $W^\pm Z$ and $W^\pm\gamma$ productions, we neglect the $b$ quark
 contribution due to the small values of the non-diagonal elements in the
 Cabibbo-Kobayashi-Maskawa matrix (CKM).

\ni
Averaging over the quark colors, a common factor of $1/3$ has to be added 
as well as a statistical factor of $1/2$ in the case of identical final
 particles ($\g\g$ and $ZZ$ productions).
\vspace{0.5cm}

\ni
Two ingredients are therefore required in order to compute cross-section
 (\ref{partons02}): the sub-process cross-section and the parton 
distributions.
As previously mentioned, a helicity amplitude approach described in Sec.4
 will be used to compute the analytic expression of sub-process 
cross-section. 
For numerical applications, Martin-Rogerts-Stirling-Distributions(MRSD)' 
structure functions are used \cite{mrsd} as they best match the recent
 data from measurements of the proton partonic content.

\vspace{0.5cm}

\ni
It is easy to deduce from Eq.(\ref{partons02}) the hadronic observable we
 want to compute.
Thanks to the unitarity breaking of anomalous couplings which leads to 
amplitudes growing with energy, the invariant mass distribution could be
 a first observable for testing TGV. 
Indeed, deviations from the SM are more apparent at higher invariant 
masses of gauge boson pair by the increase of the number of events which
 modifies the shape of the invariant mass distribution.
\vspace{0.5cm}

\ni
In the $q^{(')}_i \bar q^{(')}_j \rightarrow VV'+ X$ process, the 
inclusive differential cross-section for the production of a boson pair 
such that both intermediate bosons lie in the rapidity interval (-Y,+Y) 
is given by \cite{rendm}
\bqa
 {d\sigma\over dM}={2 M\over S} \sum_{ij}
{\int^{+Y}_{-Y}} dy_{boost} \left\{ f_i^{(a)}(x_a)f_j^{(b)}(x_b)+f_j^{(a)} 
(x_a)f_i^{(b)}(x_b) \right\} {\int^{+z_0}_{-z_0}} dz {d\hat\sigma_{ij} 
\over dz}
\label{dsigdM}
\eqa
where $z=\cost$ measures the scattering angle in the parton-parton center
 of mass (cm).

\ni
Here, it is convenient to work in terms of the rapidity of a produced 
vector boson in the hadron-hadron cm frame.
The rapidity can be decomposed in $y=y^*+y_{boost}$ where 
$\dps y^*=Tanh^{-1}(z \beta_{VV'})$ is the rapidity in the parton-parton
 center of mass frame.   
Here, we have defined
\bqa
\beta_{VV'}=\beta \biggm/ \Bigl( 1+{{m^2-m'^2}\over \hat s} \Bigr) 
\nonumber
\eqa
where $m$ and $m'$ are the masses of the $V,V'$ vectors bosons and the 
$V,V'$ velocity is
\bqa
\beta=\Bigl\lbrack {1\over \hat s^2} (\hat s+m^2-m'^2)^2 - 
{4m^2 \over \hat s} \Bigr\rbrack^{1\over 2} \ . 
\label{beta}
\eqa
The rapidity in the center of mass motion $y_{boost}$ is related to the 
parton momentum fractions $x_a$, $x_b$ via the relation 
$y_{boost}=1/2 \ \ ln(x_a/x_b)$.\\
With  $\dps \hat s=x_a x_b S =M^2$, where $M$ denotes the invariant mass
 of the vector boson pair and $S$ the $pp$ cm energy, we obtain
\bqa
x_a=\sqrt{\tau} e^{y_{boost}} \hbox{ \ and \ } 
x_b=\sqrt{\tau} e^{-y_{boost}} 
\label{jac}
\eqa
where $\dps \tau=x_ax_b=\hat s / S$.

\ni
In our numerical illustrations, we apply an experimental rapidity cut 
$\pm (Y=2)$ for the detection of vector bosons which leads to a cut in 
the parton-parton scattering angle:
$$z_0=\min \Big\lbrack Tanh (Y- |y_{boost}|)/\beta_{VV'},1\Big\rbrack \ .$$
In the case of a final photon we remove the infrared divergence of the 
$W\gamma$ pair production cross-section in restricting ourself to the 
kinematical regions of high $p_{\bot_\gamma}$ or invariant mass of the 
$W^\pm\gamma$ pair. 
This transverse momentum cut $p_{cut}$ leads to
$$ z_0<\Bigl\lbrack 1-{p^2_{cut} \over {p^2_\gamma}} 
\Bigr\rbrack^{1\over 2} 
   \mbox{ \ with \ } p_\gamma={\hat s-m^2 \over {2\sqrt{\hat s}}} \ .$$

\ni
The effects of anomalous couplings are concentrated in the region of 
small vector boson rapidity since they contribute exclusively to 
$W$, $Z$ or photon S-channel exchange.
In consequence, the transverse momentum distributions of the vector 
bosons should be particularly sensitive to the non standard WWV couplings
 especially for $W^\pm Z$ or $W^\pm\gamma$ productions which involve 
only seven anomalous couplings.

\ni
The transverse momentum distribution is easily measured by experiment 
with no ambiguity because only well measured transverse variables are 
involved. 
On the contrary, the reconstruction of the invariant mass distribution 
could be experimentally difficult, due to the unknown longitudinal 
momentum of the neutrino in the case of a final leptonic decay.
\vspace{0.5cm} 

\ni
As in the case of the invariant mass distribution, we can express from 
Eq.(\ref{partons02}) the transverse momentum distribution of the vector 
boson pair production
\bqa
\dps{ {d\sigma\over dp_\bot }= \sum_{ij}
\int \int dx_a \ dx_b \left\{ f_i^{(a)}(x_a) \ f_j^{(b)}(x_b)+
                              f_j^{(a)}(x_a) \ f_i^{(b)}(x_b) \right\} 
\ {d\hat\sigma_{ij}\over dp_\bot } }
\label{partons04}  
\eqa
with $p_\bot =\p \sint$ and $p_\parallel=\p \cost$.
After a Jacobian transformation in (\ref{partons04}) we find using 
relations (\ref{jac}):
$$
 {d\sigma\over dp_\bot } 
={1\over S} \sum_{ij} \int \int d\hat s \ dy_{boost} \
  F_{i,j}^{a,b}(\hat s,y_{boost}) \ {d\hat\sigma_{ij}\over dp_\bot } 
$$
where
$$F_{i,j}^{a,b}(\hat s,y_{boost})=
 f_i^{(a)}(\sqrt{\tau} e^{y_{boost}}) \
 f_j^{(b)}(\sqrt{\tau} e^{-y_{boost}}) 
+\mbox{} f_j^{(a)} (\sqrt{\tau} e^{y_{boost}}) \
    f_i^{(b)}(\sqrt{\tau} e^{-y_{boost}}) \ . $$
\ni
With $\cost=\pm\sqrt{ 1-\dps{p_\bot^2 / \p^2}}$, the transverse momentum 
is related to the differential cross-section by
$$
 {d\hat\sigma_{ij} \over dp_\bot }={2p_\bot  \over \beta \Delta} \ 
    {d\hat\sigma_{ij}\over d\cos\theta }
$$
where
$$\Delta=\sqrt{\hat s} \ p_\parallel
 ={1\over 2} \sqrt{\hat s^2-2\hat s(m_W^2+m_{\gamma,Z}^2+2p_\bot^2)+
(m_W^2-m_{\gamma,Z}^2)^2}$$
gives a Jacobian divergence ($\Delta=0$) for
$$\hat s_{peak}=2p_\bot^2+m_W^2+m_{\gamma,Z}^2+2\sqrt{(p_\bot^2+m_W^2)
(p_\bot^2+m_{\gamma,Z}^2)} \ .$$
Finally we find the expression of the transverse momentum distribution: 
\bqa
{d\sigma\over dp_\bot} = {1\over S} \sum_{ij} 
\int_{\hat s_{peak}}^{\hat s_{max}}d\hat s 
 \int_{-Y}^{+Y}  dy_{boost} \ 
  F_{i,j}^{a,b}(\hat s,y_{boost})
  \ {2p_\bot \over \beta \Delta} \ {d\hat\sigma_{ij}\over d\cos\theta }
\label{dsigdpt}
 \ \ .\eqa
Following relations (\ref{dsigdM}) and (\ref{dsigdpt}), both observables
 could be expressed in terms of the unpolarized differential 
cross-section. We have to calculate it for the full di-boson production
 sub-process from quark annihilation:

$$\ww \ , \ \wz \ , \ \wg  $$
and
$$\zz \ , \ \zg \ , \ \gg .$$

\ni
From Eq.(\ref{lag}), it is straightforward to calculate the explicit 
expression of the helicity amplitudes, at least at low orders in 
perturbation theory from which the unpolarized differential cross-section
 will be easily analytically derived.
\vspace{0.5cm}

\vglue 0.6cm
{\elevenbf\noindent 4. Description in term of the Helicity Amplitudes 
for Sub-Process study }
\vglue 0.2cm
\baselineskip=14pt
\elevenrm

At this level, it is convenient to fix some kinematical notations for 
the two-body parton scattering with massless incoming particles.

\ni
For the generic process $\bf q^{(')}(k) \bar q^{(')}(k') \rightarrow 
V(p)V'(p')$  where $k,k',p,p'$ denote the 4-momenta,
we have in the $VV'$ cm momentum $\dps \p=\pp={ \beta\sqrt{s}/ 2}$ where
 $\beta$ is given in Eq.(\ref{beta}).
The $s,t,u$ variables represent the usual Mandelstam variables with 
$s+t+u=m^2+m'^2$.
Following the procedure given in Ref.\cite{mery}, we will decompose the
 amplitudes in the chirality-conserving invariant form ($N_i, i=1,...,9$)
 helicity basis
\bqa
\begin{array}{lll}
\dps N_1=\ep\b '\ep.k^\prime & \dps N_2=\ep\b'\ep.k  & 
\dps N_3=\ep\b\epp.k^\prime  \\
\dps N_4=\ep\b\epp.k & \dps N_5=p\b\ep.\epp & \dps N_6=p\b'\epp.k'\ep.k\\
\dps N_7=p\b'\epp.k\ep.k'  & \dps N_8=p\b'\epp.k\ep.k  &
\dps N_9=p\b'\epp.k'\ep.k' \ .
\end{array}
\label{helici01}
\eqa
$\ep$ and $\ep'$ are the polarization vectors of the final bosons and 
could be written in the cm frame as \cite{renbook}
\bqa
\left\lbrace
\begin{array}{l}
\dps \epsilon^\mu(p,\tau=\pm1)={e^{-i\tau\varphi}\over \sqrt{2}} 
\left[\ 0 \ , \
-\tau \cost\cop-i\sip \ , \
-\tau \cost\sip+i\cop \ , \   
  \tau \su\right]\\
\dps \epsilon^\mu(p,0)= \left[ {\p\over m} \ , \
 {1\over m} e \su\cop \ , \
 {1\over m} e \su\sip \ , \
 {1\over m} e \cost \right] 
\end{array}
\right.
\nonumber
\eqa
\noindent
and
\bqa
\left\lbrace
\begin{array}{l}
\dps \epsilon'^\mu(p',\tau'=\pm1)={e^{+i\tau'\varphi}\over \sqrt{2}} 
\left[\ 0 \ , \
\tau' \cost\cop-i\sip \ , \
\tau' \cost\sip+i\cop \ , \ 
 -\tau' \su\right]\\
\dps \epsilon'^\mu(p',0)=\left[ {-|\vec{p'}|\over m'} \ , \
 {1\over m'} e' \su\cop \ , \
 {1\over m'} e' \su\sip \ , \
 {1\over m'} e' \cost\right] 
\end{array}
\right. 
\nonumber
\eqa
where 
$$\dps e= {1 \over 2\sqrt{s}} (s+m^2-m'^2) \mbox{ \ and \ } \dps 
e'= {1 \over 2\sqrt{s}} (s+m'^2-m^2) \ .$$
$\theta$ denotes the scattering angle measured in the $VV'$ rest frame 
between the incident quark momentum $\vec k$ and the final $V$ boson 
momentum $\vec p$. The angle $\varphi$ denotes the azimuthal angle and 
$\tau, \tau'=\pm 1,0$ are the polarizations of the final $V$ and $V'$ 
bosons.

\ni
Suggested by the procedure given in Ref.\cite{renee}, all the invariant 
amplitudes could be decomposed on (\ref{helici01}) as
\bqa
R=\sum_j c_j\bar v(k')N_j(a-b\g^5)u(k) \ .
\label{ampNidec}
\eqa
where a and b stand for the general vector and axial vector couplings.

\ni
The 9 helicity amplitudes for each process 
$q(\lambda) \bar q'(\lambda')\rightarrow V(\tau) V'(\tau')$, where 
$\lambda=-\lambda'=\pm 1/2$ denotes the quark helicities, are then 
directly obtained from the helicity decomposition of the $N_i$'s.
They may be written as
\bqa
 F_{\lambda\lambda'\tau\tau'}(s,\theta,\varphi)=\sum_jc_j
   F_{\lambda\lambda'\tau\tau'}(N_j,s,\theta,\varphi)(a-2b\lambda) 
\label{decheltable}
\eqa
with
$$ F_{\lambda\lambda'\tau\tau'} (N_j,s,\theta,\varphi)=
e^{i(\lambda-\lambda')\varphi}f_{\lambda\lambda'\tau\tau'}(N_j,s,\theta)$$
where
$$ f_{\lambda\lambda'\tau\tau'} (N_j,s,\theta)=
{s\lambda'\over2}\n_j\delta_{\lambda,-\lambda'} \ .$$
The expression of the $\n_i$'s are given in Table 1.\\
For each process, the helicity amplitudes will be calculated in the next
 section and are displayed in Tables 2, 3,
 4 and 5 with the following generic 
notations for the $VV'$ helicity Tables:
\def\sml{\scriptsize}
\vspace{0.2cm}
\begin{center}
\begin{tabular}{|c|c|c|c|c|c|c|}
\hline\hline
 & & {\sml$\tau=\tau'=\pm1$} & {\sml$\tau=-\tau'=\pm1$} 
& {\sml$\tau=\tau'=0$}      & {\sml$\tau=0,\tau'=\pm1$} 
& {\sml$\tau=\pm1,\tau'=0$} \\ \cline{2-7}
      & $\alpha_{VV'}$ & $\Theta_1$              & $\Theta_2$ 
              & $\Theta_3$          & $\Theta_4$          
& $\Theta_5$          \\ \hline
$\xi$ & $\Gamma_\xi$   & $\Psi_{\xi ,1}$         & $\Psi_{\xi ,2}$  
        & $\Psi_{\xi ,3}$    & $\Psi_{\xi ,4}$    & $\Psi_{\xi ,5}$\\  
\hline\hline  
\end{tabular}
\end{center}
\vspace{0.2cm}
\noindent
where $\xi$ runs in a column over the lines denoted 
$T, U, G, X, Y, Z, Z'_1, Z'_2$ and $Z'_3$.

\ni
To obtain the amplitude $F_{\lambda\lambda'\tau\tau'}$ for definite 
quark helicity $\lambda=\pm1/2$, each element $\Psi_{\xi ,n}$ of a line 
$\xi$ and column n corresponding to a definite polarization $\tau(V)$ and
 $\tau'(V')$ of the bosons has to be multiplied by the common factor 
$\Theta_n$ on top of the column and its corresponding elements 
$\Gamma_\xi$ in the second column.
The sum over all elements (i.e. over $\xi$) and the global factor 
$\alpha_{VV'}$ is to be taken.
This rule is applied for each particular final state $VV'$ 
($W^+W^-, W^\pm Z$, $W^\pm\gamma$ or $ZZ, Z\gamma, \gamma\gamma$) to
 the corresponding helicity Tables 2, 3,
 4 and 5:
\bqa
F_{\lambda\lambda'\tau\tau'}
=\alpha_{VV'} \ \Theta_n  \sum_{\xi} \Gamma_\xi \Psi_{\xi ,n} \ . 
\nonumber
\eqa
Afterwards, it is easy to compute the expression of the differential 
cross-section in terms of helicity amplitude products.
It reads
\footnote{
Bosonic states have been 
normalized as $<p'|p>=(2\pi)^32E\delta(\vec p-\vec p')$ while the 
quark states are normalized as 
 $<p'|p>=(2\pi)^3{E\over m_q}\delta(\vec p-\vec p')$. }
\bqa
\ds={\p\over16\pi s\rs}\sum_{\lambda\lambda'\tau\tau'}\vert 
F_{\lambda\lambda'\tau\tau'}\vert^2\ 
   ={\p \ \alpha_{VV'}^2\over16\pi s\rs} \sum_{n=1}^5 
\sum_{\lambda\lambda'} \vert \Theta_n  \sum_{\xi} \Gamma_\xi 
\Psi_{\xi ,n} \vert^2\ .
\label{sigmanum} 
\eqa
\ni
Each cross-section could also be expressed in the form
\bq 
\ds=C\sum_{i=1}^{N}\o_{VV'}^{\xi,\xi'}(i)\f_{VV'}^{\xi,\xi'}(i)  
\label{sigOF}
\eq
where
$\dps C=(\pi\alpha^2\p) / (4 s\rs)$ and $\dps \alpha={e^2 / 4\pi}$.
$\xi$ and $\xi'=T, U, G, X, Y, Z, Z'_1, Z'_2$ or $Z'_3$ and stand for
 the interferences of the various terms in the amplitude products.
In Eq.(\ref{sigOF}), all the $\o_{VV'}^{\xi,\xi'}(i)$ are purely 
kinematical coefficients and the $\f_{VV'}^{\xi,\xi'}(i)$ are 
combinations of coupling constants depending on the anomalous three 
boson coupling parameters.

\ni
Integrating analytically the $\o_{VV'}^{\xi,\xi'}(i)$ between $-z_0$ and
 $z_0$ it is easy to compute the integrated cross-section to obtain the
 invariant mass distribution of Eq.(\ref{dsigdM}).

\ni
The $\o_{VV'}^{\xi,\xi'}(i)$ and the $\f_{VV'}^{\xi,\xi'}(i)$ depend 
strongly on the final bosonic state and will be given below for each 
process.
We divide them into three sets, the first set which already exists in 
the standard case, the second set deals with couplings conserving $CP$
 and the last set appears only through $CP$-violating ones.
Both formulations of the cross-section in Eq.(\ref{sigmanum}) and  
Eq.(\ref{sigOF}) give two different numerical calculations tools and 
allow for various cross-checks.
\vspace{0.5cm}

\vglue 0.6cm
{\elevenbf\noindent 5. $\ww$ process}
\vglue 0.2cm
\baselineskip=14pt
\elevenrm

The three lowest order Feynman diagrams for $W^+ W^-$ gauge boson pair
 production from the $q^{(')}\bar q^{(')}$ annihilation are shown in
 Fig.2 and are very similar to the $\ee$ contributions.
The U- and T-channels correspond respectively to the up or down 
quark-antiquark annihilation.
The S-channel induces a sensitivity to both $WW\gamma$ and $WWZ$ 
couplings.
\vspace{0.5cm}

\ni
The quark exchange part in the T-channel is written as 
$$ R_{WW}(t)=-{e^2\over t}\vb(k') \ep\b'(k\b-p\b)\ep\b a_T(1-\g^5) u(k) $$
where $a_T=2a_{Wii}^2$ correspond to the general vector and axial vector
 couplings $a_{Wij}$ and $b_{Wij}$ for $q_i\bar{q_j}^\prime W^\pm$ vertex:
$$ a_{Wij}=b_{Wij}={1\over 2\sqrt{2} \sin\tw}V_{ij} \ . $$
$V_{ij}$ are the elements of the CKM quark mixing matrix and $\tw$ the
 weak mixing angle of Weinberg.

\ni
According to Eq.(\ref{ampNidec}), the decomposition on the helicity 
basis (\ref{helici01}) gives
\bqa
R_{WW}(t) &=&-{e^2 \over st}\vb(k') \Big\lbrace
        (s+\mw-t)(N_2-N_3)+(\mw-t)(N_1-N_4)\nonumber\\
     &+& s N_5 +2(N_7-N_6) \Big\rbrace a_T(1-\g^5) u(k) \ . 
\label{amptww}
\eqa
In the same way the U-channel is written as 
\bqa
R_{WW}(u)&=&-{e^2\over u}\vb(k') \ep\b(k\b-p\b')\ep\b' a_U(1-\g^5) u(k) 
\nonumber\\
         &=&-{e^2 \over su}\vb(k') \Big\lbrace
         (s+\mw-u) (N_4-N_1) +(\mw-u) (N_3-N_2) \nonumber\\
         & & - s N_5 +2(N_7-N_6) \Big\rbrace a_U(1-\g^5) u(k)
\label{ampuww}  
\eqa
with $a_U=a_T$.
\ni
For the contribution of the S-channel, we use the notations of 
Fig.3. 
It is straightforward to derive the Feynman rules for the three-boson 
vertices in $W^+W^-$ production from the phenomenological effective 
Lagrangian (\ref{lag}):
\bqa
\v^V&=&g_V[\ep.\epp(p-p^\prime)_\mu
-2\epp.p\ep_\mu+2\ep.p^\prime\epp_\mu]
            +x_V[\ep.p'\ep'_\mu-\epp.p\ep_\mu]  \nonumber\\
& &\mbox{}+{y_V\over\mw}[q_\nu g_{\lambda\mu}-q_\lambda
g_{\nu\mu}][p^{\prime\lambda}\ep^{\prime\rho}
-p^{\prime\rho}\ep^{\prime\lambda}][p_\rho\ep^\nu-p^\nu\ep_\rho]
   \nonumber\\
& &\mbox{} +i{z_V\over\mw}
\eps q^\nu(p^\prime-p)^\rho [ \epp.p \ep^\sigma - 
\ep.p^\prime \ep^{\prime\sigma} ]
+iz'_{1V} [ \epp.p \ep_\mu + \ep.p^\prime \epp_\mu]
\nonumber\\
& &\mbox{ } +z'_{2V}\eps q^\nu\ep^\rho\ep^{\prime\sigma}
 -{z'_{3V}\over\mw}(p^\prime-p)_\mu\epst
q^\tau(p^\prime-p)^\nu\ep^\rho\ep^{\prime\sigma}\  
.\label{wwfeynrules}
\eqa
The two amplitudes in the S-channel ($V=\gamma$ or $Z$) are    
\bqa
R^V_{WW}(s)={e^2\over D_V(s)}\bar v(k^\prime)\g^\mu
(a_{Vi}-b_{Vi}\g^5)u(k)\v^V\ 
\nonumber
\eqa
where $D_\gamma(s)=s$ and, in the $s>4m_Z^2$ case, the $Z$ propagators
 is approximated to $D_Z(s)\simeq s-m^2_Z$ .

\ni
The vector boson-$q\bar q$ couplings $a_{Vi}$ and $b_{Vi}$ are kept in
 standard forms: for $q_iq'_i\gamma$ vertex
\bqa
a_{\gamma i}=Q_i \mbox{ , } b_{\gamma i}=0       
\nonumber
\eqa
and for $q_iq'_iZ$ vertex
\bqa
a_{Z i}={1\over 4\sin\tw\cos\tw}(\tau_3^i-4Q_i\sin^2\tw) \mbox{ \ , \ }
b_{Z i}={1\over 4\sin\tw\cos\tw}\tau_3^i        
\nonumber
\eqa
where $Q_i$ and $\tau_3^i$ are the electric charge and weak isospin 
projection of the i-quark:
\bqa
\left\lbrace
\begin{array}{l}
   Q_i=\;\;\; {2 / 3}\hbox{ , } \tau_3^i=\;\;\; 1 \hbox{ for up quark}\\
   Q_i=  -    {1 / 3}\hbox{ , } \tau_3^i=  -    1 \hbox{ for down quark.}
\nonumber
\end{array}
\right.
\eqa
Following Ref.\cite{renee}, the decomposition of Eq.(\ref{wwfeynrules})
 on helicity basis (\ref{helici01}) leads to: 
\bqa
R^V_{WW}(s)& =&{e^2\over D_V(s)}\bar
v(k') \Big\{ 2g_V(N_1+N_2-N_3-N_4+N_5)  \nonumber \\
&+ &\mbox{} x_V(N_1+N_2-N_3-N_4)\nonumber \\
&+ &\mbox{} {y_V\over\mw} \Big[ sN_5+\mw(N_1+N_2-N_3-N_4) \nonumber \\
& &\mbox{}+2(N_6+N_7+N_8+N_9) \Big] \nonumber \\
&- &\mbox{} {z_V\over\mw} \Big[ (t-u)(N_1+N_2-N_3-N_4)+4(N_6-N_7) \Big]
 \g^5
\nonumber \\
&+ &\mbox{} iz'_{1V}(N_1+N_2+N_3+N_4)+iz'_{2V}(N_1+N_4-N_2-N_3)\g^5 
\nonumber\\
&- &\mbox{} i{z'_{3V}\over\mw} \Big[ (s-4\mw)(N_1-N_2-N_3+N_4)-2(t-\mw+
{s\over 2})
\times \nonumber \\
& &  (N_1+N_2+N_3+N_4)+4(N_8-N_9) \Big] \g^5 \Big\} 
(a_{Vi}-b_{Vi}\g^5) u(k)\
.\label{ampsww}
\eqa

\ni
According to the procedure given in Sec.4 and to Eq.(\ref{decheltable}),
 the Eq.(\ref{amptww}), (\ref{ampuww}) and (\ref{ampsww}) lead to the
 helicity Table 2 for $\ww$.

\vspace{0.5cm}\ni
Defining for convenience
$\dps \chiz=s / (s-\mz)$ , $a_Z=a_{Zi}$ , $b_Z=b_{Zi}$ , 
$a_T=a_U=2a_{Wij}^2$ 
and the intermediate functions as follows:
\bqa
 \G_{WW}^{V,V}(c_\gamma,c_\gamma^\prime,c_Z,c_Z^\prime) &=&
       c_\gamma c_\gamma^\prime a_\gamma^2 +
       c_Z c_Z^\prime (a_Z^2+b_Z^2) \chiz^2 \nonumber\\
       &+& (c_\gamma c_Z^\prime+c_Z c_\gamma^\prime) a_\gamma a_Z \chiz  
\nonumber\\
 \G_{WW}^{V,(T\ or\ U)}(c_\gamma,c_Z) &=& a_T \Big[ a_\gamma c_\gamma+
(a_Z+b_Z) \chiz c_Z \Big]  
\nonumber\\
 \G_{WW}^{Z,Z} &=& z_\gamma^2 a_\gamma^2 + z_Z ^2 (a_Z^2+b_Z^2) \chiz^2 +
      2 z_\gamma z_Z  a_\gamma a_Z \chiz  \nonumber\\
 \G_{WW}^{GXY,Z} &=& a_\gamma b_Z \chiz \Big[ z_\gamma (2 g_Z+x_Z+y_Z)+z_Z 
      (2 g_\gamma+x_\gamma+y_\gamma) \Big] \nonumber\\
         &+& 2 a_Z b_Z \chiz^2 z_Z  (2 g_Z+x_Z+y_Z)  \nonumber\\
 \G_{WW}^{Z,(T\ or\ U)} &=&-a_T \Big[ a_\gamma z_\gamma+(a_Z+b_Z) \chiz
 z_Z \Big] 
\nonumber
\eqa
\noindent
we find with the help of helicity Table 2 the expressions
 of the $\f_{WW}^{\xi,\xi'}(i)$ and $\o_{WW}^{\xi,\xi'}(i)$ terms of
 Eq.(\ref{sigOF}) for $\ww$.

\ni
These terms are summarized in Tables 6, 7 and 8.
\vspace{0.5cm}

\vglue 0.6cm
{\elevenbf\noindent 6. $\wz$ and $\wg$ processes}
\vglue 0.2cm
\baselineskip=14pt
\elevenrm

The three lowest order Feynman diagrams for $W^\pm V$, $V=\gamma,Z$ 
production from $q\bar q'$ annihilation are shown in Fig.4.

\ni
For $W^+V$ production, the T-channel amplitude is
\bqa
R_{W^+ V}(t)=- {e^2\over t}\vb(k') \ep\b'(k\b-p\b)\ep\b a_T(1-\g^5) u(k)
 \nonumber
\eqa
with $a_T=a_{Wij}(a_{Vj}+b_{Vj})$. The decomposition on the helicity
 basis (\ref{helici01}) gives
\bqa
R_{W^+ V} (t) &=&- {e^2 \over st}\vb(k') \Big\lbrace 
         -(s+\mv-t) N_3 + (s+\mw-t) N_2 +(\mv-t)N_1 \nonumber\\
 &-& (\mw-t)N_4 + s N_5 +2(N_7-N_6)   \Big\rbrace a_T(1-\g^5) u(k)\ . 
\label{amptwv}
\eqa
In the same way, the contribution of the U-channel amplitude is
$$R_{W^+ V}(u)=- {e^2\over u}\vb(k') \ep\b(k\b-p\b')\ep\b' a_U(1-\g^5)
 u(k)$$
with $a_U=a_{Wij}(a_{Vi}+b_{Vi})$ and leads through (\ref{helici01}) to
\bqa
R_{W^+ V} (u) &=&- {e^2 \over su}\vb(k') \Big\lbrace
         (s+\mv-u) N_4 - (s+\mw-u) N_1 +(\mw-u)N_3 \nonumber\\
 &-& (\mv-u)N_2  - s N_5 +2(N_7-N_6)  \Big\rbrace a_U(1-\g^5) u(k) \ .
\label{ampuwv}
\eqa

\ni
With the notations of Fig.5, we derive the Feynman rules
 for the $W^\pm (Z,\gamma)$ production with the complete TGV contribution
 of Lagrangian (\ref{lag}):
\bqa
\v^V&=&
 \eta g_V \Big[ \ep.\epp(p-p^\prime)_\mu
               -2\epp.p\ep_\mu+2\ep.p^\prime\epp_\mu \Big]
+\eta x_V \Big[ \ep.\ep' p_\mu-\epp.p\ep_\mu \Big]  \nonumber\\
& &\mbox{}-\eta {y_V\over\mw}
         \Big[ \ep.p^\prime \ep'.p (p-p^\prime)_\mu 
         - p^2 \ep.p^\prime \ep'_\mu + p^{\prime 2} \ep'.p \ep_\mu
         - \ep.\ep' p^\prime.q p_\mu + \ep.\ep' p.q p^\prime_\mu \Big]
\nonumber\\
& &\mbox{}-i{z_V\over\mw} \epsilon_{\rho\sigma\alpha\beta}
p^\alpha \ep^\beta (2p^\prime+p)^\rho(\ep'.p g^{\mu\sigma}-\ep'^\sigma
 p^\mu)
\nonumber\\
& &\mbox{} -iz'_{1V} \Big[ \ep.\ep'p_\mu + \ep'.p \ep_\mu \Big]
-\eta z'_{2V}\eps p^\rho \ep^\sigma \ep'^\nu
\nonumber\\
& &\mbox{ }
-\eta {2 z'_{3V} \over \mw} \eps p^\rho \ep^{\prime\nu} (2p'+p)^\sigma 
\ep.p'
\nonumber
\eqa
where $\eta=\mp 1$ for a $W^\pm$ in the final state.\\
The $W^+$ formation part is given by
\bqa
R_{W^+V}(s)={e^2\over D_W(s)}\bar v(k^\prime)\g^\mu
             (a_{Wij}-b_{Wij}\g^5)u(k) \ \v^V\ 
\label{swV06}
\eqa
where the $W$ propagator is approximated as $D_W(s)\simeq s-m_W^2$.\\
\ni
The decomposition of Eq.(\ref{swV06}) on the helicity basis
 (\ref{helici01}) gives
\begin{eqnarray}
R_{W^+ V}(s)&=&{e^2\over D_W(s)}\bar v(k') \Big\{
-2 g_V(N_1+N_2-N_3-N_4+N_5) \nonumber \\
&-&\mbox{ } x_V (N_5-N_3-N_4)\nonumber \\
&-&\mbox{ }  {y_V\over\mw} \Big[ sN_5+\mv(N_1+N_2)-\mw(N_3+N_4) 
\nonumber \\
& &\mbox{}+2(N_6+N_7+N_8+N_9) \Big] \nonumber \\
&+ &\mbox{ }{2 z_V\over\ s \mw} \Big[ (N_6-N_7)(s+\mv-\mw)
          -{s^2\bs\over4}(N_1-N_2+N_4-N_3)\nonumber \\
& &\mbox{} 
          +{1\over4}s\beta\cost(s+\mv-\mw)(N_1+N_2-N_3-N_4) \Big] \g^5 
 \nonumber \\
&- &\mbox{ }iz'_{1V}(N_3+N_4+N_5) \nonumber \\
&+ &\mbox{ } i z'_{2V} \Big[ {1\over 2s}(s+\mv-\mw)(N_1-N_2-N_3+N_4) 
\nonumber \\
& &\mbox{}           -{1\over 2}\beta\cost(N_1+N_2-N_3-N_4)
                     -{2\over s}(N_6-N_7) \Big] \g^5 \nonumber\\
&+ &\mbox{ } i {z'_{3V}\over \mw} \Big[ 4(N_8-N_6+N_7-N_9)    \nonumber\\
& &\mbox{}           -2s\beta\cost(N_1+N_2) \Big] \g^5 \Big\}
                      a_{Wij}(1-\g^5)u(k) \ .
\label{ampswv}
\end{eqnarray}
Eq.(\ref{amptwv}), (\ref{ampuwv}) and (\ref{ampswv}) are similar for
 $W^- V$ production and according to the procedure given in Sec.4 and to
 Eq.(\ref{decheltable}), lead to the helicity Table 3 for
 $W^\pm Z$ final state and Table 4 for $W^\pm\gamma$
 final state.
\vspace{0.5cm}

\ni
As before, we can find the expressions of the 
$\f_{W(Z\ or\ \gamma)}^{\xi,\xi'}(i)$ and $\o_{W(Z\ 
or\ \gamma)}^{\xi,\xi'}(i)$ terms of Eq.(\ref{sigOF}) for 
$\wz$ and $\wg$ with respectively the helicity Tables 3
 and 4.\\
Defining for $W^\pm Z$ and $W^\pm\gamma$ productions 
$\dps \chiw=1/(s-\mw)$ and the intermediate functions as follows:
\bqa
 \G_{WV}^{V,V}(c,c') &=&  c c' a_{Wij}^2 \chiw^2\nonumber\\
 \G_{WV}^{V,T}(c)    &=& a_{Wij}^2 (\avj+\bvj) \chiw c \nonumber\\
 \G_{WV}^{V,U}(c)    &=& a_{Wij}^2 (\avi+\bvi) \chiw c \nonumber\\
 \G_{WV}^{GXY,Z}     &=& -a_{Wij}^2 \chiw^2 z_V (2 g_V+ x_V + y_V)
\nonumber\\
 \G_{WV}^{Z,T}       &=&- a_{Wij}^2 (\avj+\bvj) \chiw z_V\nonumber\\
 \G_{WV}^{Z,U}       &=&- a_{Wij}^2 (\avi+\bvi) \chiw z_V\nonumber\\
 \G_{WV}^{Z,Z}       &=& \G_{WV}^{VV}(z_V,z_V)\nonumber\\
 \G_{WV}^{T,T}       &=& a_{Wij}^2 (\avj+\bvj)^2\nonumber\\
 \G_{WV}^{U,U}       &=& a_{Wij}^2 (\avi+\bvi)^2\nonumber\\
 \G_{WV}^{T,U}       &=& a_{Wij}^2 (\avi+\bvi)
             (\avj+\bvj)\nonumber
\eqa
\noindent
we find the results of Table 9 for the 
$\f_{WV}^{\xi,\xi'}(i)$ where $V=Z$ or $\gamma$. 
The kinematical variables $\o_{WV}^{\xi,\xi'}(i)$ are summarized in 
Tables 10 and 11 for $W^\pm Z$ production and 
in Tables 12 and 13 for $W^\pm \gamma$ production.
\vspace{0.5cm}

\vglue 0.6cm
{\elevenbf\noindent 7. Discussion }
\vglue 0.2cm
\baselineskip=14pt
\elevenrm

As previously mentioned, in the most general case we have to deal with
 seven $WW\gamma$ and seven $WWZ$ couplings.
When we allow for more than one anomalous coupling to be non zero it is
 clear that the possibilities of correlations can not be excluded. 
The question of making a significant test of the SM involves three steps: 
1) measuring independently and precisely each $WWV$ vertex; 
2) checking whether they agree with the SM values;  
3) possibly, disentangling the effects of the various anomalous couplings.
\vspace{0.5cm}

Eq.(\ref{efflag}) leads generally to a $(\sqrt{\hat s}/\Lambda)^{d-4}$
 energetic behavior \cite{rendyn}.
For an expected value of the NP scale $\Lambda$ in the TeV range, it seems
 obvious that one does not expect a violent low energy departure from the
 standard predictions,
i.e. the 14 parameters $\delta_V,x_V,y_V,z_V,z'_{1V},z'_{2V}$ and 
$z'_{3V}$ $(V =\g,Z)$ should have reasonably small values.
Nevertheless, for large values of the di-boson invariant mass 
$\sqrt{\hat s}$ and provided that the signal is not overwhelmed by the
 background, the non-standard contributions to the helicity amplitudes
 would dominate. 
Information on anomalous $WWV$ couplings should be obtained in comparing
 the shapes of the measured and predicted transverse momenta or invariant
  mass distributions for each $W^\pm V$, ($V=W^\pm, Z, \gamma$)
 productions.
As suggested in Ref.\cite{baur} and \cite{ratio}, the uncertainty on the
 structure functions should be reduced by considering ratios of 
non-standard and standard contributions as for example: 

$${\sigma(pp\rightarrow W^\pm Z) \ Br(W^\pm \rightarrow l^\pm \nu)
 \ Br(Z \rightarrow  l^+ l^- )
\over \sigma(pp\rightarrow ZZ) \ Br(Z \rightarrow  l^+ l^- )^2} \ .$$

\vspace{0.3cm}\ni
The presence of anomalous couplings in the $WWV$ vertex will yield an 
enhanced number of events at large invariant mass $M_{WV}$ or $p_{\bot_V}$
 transverse momentum of the $W^\pm V$ system.
As an illustration, the sensitivity of the various distributions to 
deviations from standard couplings has been tested by changing the 
parameters from their SM expectations. 
The resulting shapes are displayed for $V=\gamma , Z$ and for several 
choices of anomalous couplings in Fig.6 and Fig.7.
The leptonic branching ratios per lepton species are included in the 
cross-section and have been taken to be 
$Br(W^\pm Z \rightarrow       l_1^\pm \nu_1  \  l_2^+ l_2^- )=0.36$\% and 
$Br(W^\pm \gamma \rightarrow  l_{1,2}^\pm \nu_{1,2}   \  \gamma)=10.7$\%
 with $l^\pm_{1,2}=e^\pm$ or $\mu^\pm$.
Note that around $300$ GeV for $p_{\bot_Z}$ and 150 GeV for 
$p_{\bot_\gamma}$, one can get an effect between $5$ and $50$ \% even 
for moderate values of the anomalous couplings. 
Of course, for the  same values of the couplings, higher effects can be
 reach at higher $p_{\bot_{(Z,\gamma)}}$.

\ni
Here, as a starting point, we do not allow for several parameters to be 
simultaneously non vanishing and only one coupling is assumed to deviate
 from the SM at a time.

\ni
In practice, large cancellations between the different terms cannot be
 excluded and the full set of parameters have to be considerate 
simultaneously.
In this case, the limits of observability for the couplings will be 
obtained by a multi-parameter analysis.
To perform this analysis, we can use a Maximum Likelihood fit 
\cite{MLfit} or methods based on density matrix or optimal observables
 respectively exposed in Ref.\cite{renee} and \cite{OOmet}.  
The helicity Tables 2,3,4
 and the analytic expressions of Tables 6 to 13 
will be useful to perform this analysis which has to be done in a phase
 space region where the effects of non standard Three Vector Boson 
Couplings are much larger than the background. 

\ni
For completeness, we deduce the helicity amplitudes of the standard 
$ZZ,\ Z\gamma$ and $\gamma\gamma$ productions from the previous $W^+W^-$ 
and $W^\pm V$ cases.
We reproduce them in Table 5.
Only the T- and U-channels are allowed in our scenario and using the same
 notations as for the other productions, they lead to the 
$\o_{VV'}^{\xi,\xi'}(i)$ and $\f_{VV'}^{\xi,\xi'}(i)$ expressions of 
Table 14 (with $VV'=ZZ \ ,\ Z\gamma \mbox{ or } 
\gamma\gamma $).

\vspace{0.5cm}\ni
Now, we would like to show why, starting from the most general case of 14
 free parameters, one can in principle construct a strategy for 
disentangling the parameters if one identifies the Longitudinal($L$) or
 Transverse($T$) polarization states of the final $W^{\pm}$ and $V$ 
bosons.

\ni
In what follows, we shall discuss the contributions of anomalous 
couplings to the helicity amplitudes.
\ni
The helicity Table 2 corresponding to $W^+W^-$ production
 having an identical structure to the one done for the $\ee$ process, 
the reader is referred to Ref.\cite{renee} and \cite{lep} for extensive
 discussion.
A glance at the helicity amplitudes of Tables 3 and 
4 allows us immediately to draw several useful 
conclusions on the contributions of the various couplings to the 
$W^\pm Z$ and $W^\pm \gamma$ final states.

\vspace{0.5cm}
For $W^\pm Z$ production, the helicity amplitudes lead to strong mixing 
among the three types of C- and P-conserving forms corresponding to the 
$\delta_Z,\ x_Z$ and $y_Z$ couplings (i.e. the deviation of Yang-Mills
 coupling, the magnetic and the quadrupole terms). 
However, these three forms contribute, like the CP-conserving but 
C- and P- violating coupling $z_Z$ and like the three other CP-violating
 couplings $z'_{1Z}, \ z'_{2Z}$ and $z'_{3Z}$, to the final polarization
 states ($TT$, $LL$ and $LT$) with very different weights. 
Thus, the measurement of the vector boson polarization (spin density
 matrix elements) will be crucial for separating their effects.
More precisely, as shown in Table 3, one observes that 
$\delta_Z$, $x_Z$, and $z'_{1Z}$ contribute to all of the $TT$, $LL$ and
 $LT$) state while $y_Z$, $z_Z$, $z'_{2Z}$ and $z'_{3Z}$ contribute to
 $TT$ and $LT$ but not to $LL$ state.
Similarly to $W^+W^-$ production, there is no $TT$ contribution coming
 from the S-channel in the case of opposite polarization, 
$\tau=-\tau'=\pm 1$.

\ni
The standard and all anomalous contributions to the $Z$ transverse 
momentum distribution with $TT,LL$ or $LT$ final $W^\pm Z$ polarization
 states are shown in Fig.8.
Due to the same dependence in the $q\bar{q}$ sub-process, an identical 
behavior can be observed in the shapes of the invariant mass distribution
 of the $W^\pm Z$ pair.
For conciseness, these shapes are not included in this study and are
 postponed in a forthcoming paper.

\ni
According to the helicity Table 3, we observe in Fig.8.b
 that for the $\delta_Z$ coupling, the high energy behavior is dominated 
by $TT$ and $LL$ productions whereas, as shown in Fig.8.c, the $x_Z$
 coupling contribute essentially to $TT$ and $LT$ productions.
The amplitudes corresponding to $y_Z$ (see Fig.8.d) and $z'_{1Z}$ 
(see Fig.8.f) couplings are respectively dominated by $TT$ and $LL$ 
productions.
Furthermore, the $z_Z$, $z'_{2Z}$ and $z'_{3Z}$ couplings contribute
 essentially to $LT$ final state but with different energy dependence 
for this production (Fig.8.e, Fig.8.g and Fig.8.h).

\ni
All those properties should allow a clear separation between all the 
different $WWZ$ anomalous couplings.

\vspace{0.5cm}
The $W^\pm \gamma$ final state leads to helicity amplitudes very similar 
to the previous $W^\pm Z$ process except for the $z'_{3V}$ contribution 
which vanishes in $W^\pm\gamma$ production.
In 
Fig.9.a, 9.b, 9.c and 9.d, we show respectively the SM and the $x_\gamma$,
 $y_\gamma$, $z_\gamma$ contributions to the photon transverse momentum 
distribution with $TT$ or $LT$ final $W^\pm\gamma$ polarization states.

\ni
As in the $W^\pm Z$ case, one observes the same behavior between the 
invariant mass distribution and the photon transverse momentum 
distribution for all allowed final polarizations.
Helicity Table 4 shows that the $\delta_\gamma$ and 
$x_\gamma$ couplings have the same contribution.
Their high energy behavior is dominated by the $1/m_W \sqrt{s}$ term of
 the $LT$ state as for $z'_{1\gamma}$ and $z'_{2\gamma}$ couplings
 (see Fig.9.b).
Thus, there is no possibility to disentangle any of these couplings
 through their effect in $W^{\pm}\gamma$ production only.
A complete separation of the effects for these $WW\gamma$ anomalous 
couplings needs an additional analysis.
Using a multiparameter fit, we can study for example the interplay 
between different anomalous couplings being simultaneously non-zero in
 $W^\pm \gamma$ production.
We can also test the $W^+W^-$ production which is sensitive to both 
$WWZ$ an $WW\gamma$ anomalous couplings and where the previous 
contributions are not identical.

\ni
The $y_\gamma$ and the anapole coupling $z_\gamma$ can be separated from
 the others as the $TT$ state has a strong increase with $\sqrt{s}$
 for $y_\gamma$ (Fig.9.c), while the contribution of the $z_\gamma$ term 
is dominated by the $LT$ contribution (Fig.9.d).   
%

\newpage

\vglue 0.6cm
{\elevenbf\noindent 8. Conclusions}
\vglue 0.2cm
\baselineskip=14pt
\elevenrm

As remembered in Sec.1, the investigation of indirect signal of New 
Physics results of many strong theoretical motivations.
The actual precision tests on Boson-Fermion interactions being in 
agreement with the standard predictions with a high level of accuracy, 
we have chosen to probe the existence of NP through the precise and 
direct test of the bosonic self interactions which should be feasible at
 the next generation of hadron colliders.
Leaving the four boson couplings for a separate analysis, we restricted 
ourself to the test of the Three Gauge Boson couplings predicted by the
 Standard Model.

\ni
In this paper we have given several calculation tools which should be 
useful to probe the most general structure of the $WWV$ ($V=\gamma$ or
 $Z$) vertex in a model independent way through the precise measurement
 of the $W^+W^-, W^\pm Z$ and $W^\pm \gamma$ bosonic pair production.

\ni
In Sec.5 and 6, we have computed the helicity amplitudes for the 
production of all TGV-sensitive two boson final states via 
quark-antiquark annihilation.
We deduced from these Tables the full analytic expressions of the tree 
level Cross-Sections for all productions and in function of the 14 free
 parameters of the most general $WWV$ $(V=\gamma, Z$) effective 
Lagrangian.
These Cross-Sections should be very useful to predict the sensitivity 
of present and future proton-(anti)proton colliders to the complete set
 of anomalous trilinear couplings given in Sec.2.
We exhibit in Sec.8 the sensitivity of the invariant mass and the 
transverse momentum distributions studied in Sec.3 and 7 to the various
 anomalous couplings for $W^\pm Z$ and $W^\pm \gamma$ productions.
Some other hadronic observables as the amplitude zeros, the rapidity 
correlations and cross section ratios should also be sensitive 
\cite{baur,ampzero,rendm} and can be tested in performing a 
model-independent multi-parameter analysis of real or simulated data
 through fits and shapes of contour plots.
In the eventuality where a significant departure from the standard 
expectations is measured, the exploration of the underlying NP through 
bosonic sector will be indissociable from an identification of the 
anomalous couplings responsible for this deviation.
This goes unavoidably through the disentangling of the 14 free parameters
 effects of the full effective TGV Lagrangian (\ref{lag}).

\ni
As anomalous couplings contribute to helicity amplitudes given in 
Tables 2 to 4 with different weights, we 
show in Sec.7 that the identification of the Longitudinal or Transverse
 polarization of both final bosons will be very useful to separate those
 effects.

\ni
In conclusion, we find that if one identifies the polarization state of
 the final $W$ and $Z$ bosons and in the case where only one anomalous
 coupling is supposed to be non-zero at a time, the helicity 
Table 3 should allow us to identify and separate the
 contributions of all the seven $WWZ$ anomalous couplings  
$\delta_Z, \ x_Z, \ y_Z, \ z_Z, \ z'_{1Z}, \ z'_{2Z}, \ z'_{3Z}$, 
through the analysis of the $W^\pm Z$ production only.

\ni
On the contrary, given the helicity Table 4, only the 
existence of non-zero $\delta_\gamma$, $x_\gamma$, $z'_{1\gamma}$ or 
$z'_{2\gamma}$ couplings can be tested through single $W^\pm \gamma$ 
production without possibility to identify from which coupling came the
 deviation.
Nevertheless, the $y_\gamma$ and $z_\gamma$ couplings should be clearly
 separated from the others through an analysis of the final $W^\pm$ 
polarization states in $W^\pm\gamma$ process only.
A complete separation of the effects of the other $WW\gamma$ anomalous
 couplings needs an additional analysis.

\ni
These properties which allow a clear separation of the different types of
 $WWV, (V=Z \mbox{ or } \gamma)$ couplings can only be achieved if one
 has enough luminosity for providing a large number of bosonic pair 
events decaying both in leptonic final state. 
Due to the high luminosity expected at LHC and the future luminosity 
upgrade at FERMILAB, this condition should be realized in the next 10 
years at least. 

\ni

Finally, we stress that the analytic expressions of the helicity Tables
 and Cross-Section given in this paper should be very suitable for a 
general model independent study of the sensitivity of actual and future
 measurements to an indirect signal of NP in the bosonic sector
\footnote{
All relevant analytic expressions are available upon
 request by E.Mail at: NUSS@lpm.univ-montp2.fr}
.
%
\vglue 0.6cm
{\elevenbf\noindent 9. Acknowledgments}
\vglue 0.2cm
\baselineskip=14pt
\elevenrm

\ni
I would like to thank J.L. Kneur for his very helpful suggestions in 
writing the Fortran codes.
I am especially grateful to J. Layssac, F.M. Renard and G. Moultaka for
 many valuable and stimulating discussions and a critical reading of the
 manuscript.

\newpage
\vglue 0.6cm
{\elevenbf\noindent 9. References}
\vglue 0.2cm
\baselineskip=14pt
\elevenrm
%

%
\newpage
\pagestyle{empty}
\ni
{\huge List of Tables}
\vspace{0.5cm}

\begin{tabular}{ll}
Table 1:  &  Decomposition of the helicity basis.\\
Table 2:  &  Helicity Table for $W^+W^-$ production.\\
Table 3:  &  Helicity Table for $W^\pm Z$ production. 
$\eta=\mp 1$ for $W^\pm$ production.\\
Table 4:  &  Helicity Table for $W^\pm\gamma$ production. 
$\eta=\mp 1$ for $W^\pm$ production.\\
Table 5:  &  Helicity Tables for $ZZ,Z\gamma$ and $\gamma\gamma$ 
production.\\
Table 6:  &  $\f_{WW}^{\xi,\xi'}(i)$ coefficients for 
$W^+W^-$ production.\\
Table 7:  &  Standard $\o_{WW}^{\xi,\xi'}(i)$ coefficients for 
$W^+W^-$ production.\\
Table 8:  &  Non standard $\o_{WW}^{\xi,\xi'}(i)$ coefficients for 
$W^+W^-$ production.\\
Table 9:  &  $\f_{WV}^{\xi,\xi'}(i)$ coefficients for $W^\pm Z$ and 
$W^\pm\gamma$ production.\\
Table 10: &  Standard $\o_{WZ}^{\xi,\xi'}(i)$ coefficients for 
$W^\pm Z$ production.\\
Table 11: &  Non standard $\o_{WZ}^{\xi,\xi'}(i)$ coefficients for 
$W^\pm Z$ production.\\
Table 12: &  Standard $\o_{W\gamma}^{\xi,\xi'}(i)$ coefficients for
 $W^\pm \gamma$ production.\\
Table 13: &  Non standard $\o_{W\gamma}^{\xi,\xi'}(i)$ coefficients
 for $W^\pm \gamma$ production.\\
Table 14: &  Standard $\o_{VV'}^{\xi,\xi'}(i)$ and $
\f_{VV'}^{\xi,\xi'}(i)$ coefficients for \\
          &  ${\bf q^{(')}_i \bar{q}^{(')}_i  \rightarrow VV'}$  
with $VV'=ZZ,Z\gamma,\gamma\gamma$.
\end{tabular}

\vspace{2cm}

\ni
{\huge List of Figures}
\vspace{0.5cm}

\begin{tabular}{ll}
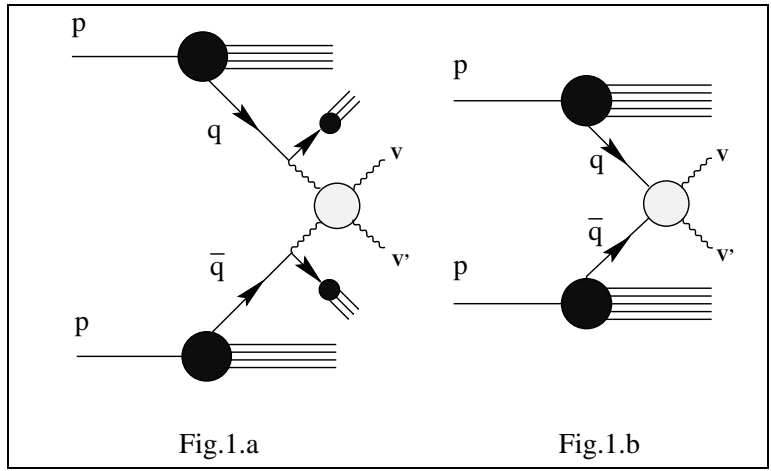
Fig.1 &  Bosonic pair production via Bosons fusion (a) 
and quarks annihilation (b).\\
Fig.2 &  Feynman graphs for $W^+W^-$ production.\\
Fig.3 &  $W^+W^-V$ vertex.\\
Fig.4 &  Feynman graphs for $W^\pm Z$ and $W^\pm\gamma$ productions.\\
Fig.5 &  $W^\pm W^\pm V$ vertex.\\
Fig.6 &  Invariant Mass distributions for $W^\pm Z$ (Fig.6.a) and 
$W^\pm \gamma$ (Fig.6.b)\\
      &  productions with various non-zero anomalous couplings.\\
Fig.7 &  Transverse Momentum distributions for $W^\pm Z$ (Fig.7.a) and \\ 
      &  $W^\pm \gamma$ (Fig.7.b) productions with various non-zero 
anomalous couplings.\\
Fig.8 &  Longitudinal($L$) and Transverse($T$) final state polarizations 
in \\
      &  Standard Momentum distribution (Fig.8.a) and with\\
      &  anomalous couplings for $W^\pm Z$ production: $\delta_Z$ 
(Fig.8.b) ,\\
      &  $x_Z$ (Fig.8.c), $y_Z$ (Fig.8.d), $z_Z$ (Fig.8.e) and 
$z'_{1Z}$ (Fig.8.f)\\
      &  $z'_{2Z}$ (Fig.8.g) and $z'_{3Z}$ (Fig.8.h).\\
Fig.9 &  Longitudinal($L$) and Transverse($T$) final state polarizations 
for  \\
      &  $W^\pm \gamma$ production in Standard Momentum distribution 
(Fig.9.a) and \\
      &  with anomalous couplings: $x_\gamma$ (Fig.9.b), $y_\gamma$ 
(Fig.9.c) and $z_\gamma$ (Fig.9.d).
\end{tabular}

%
\end{document}